\documentclass[superscriptaddress,amsmath,amssymb,prl,nofootinbib,twocolumn]{revtex4-2}
\usepackage{hyperref}
\usepackage[graphicx]{realboxes}
\usepackage{placeins} 
\usepackage{babel}
\usepackage{times}
\usepackage[usenames,dvipsnames]{color}
\usepackage{soul}
\usepackage[utf8]{inputenc}
\usepackage{float}
\usepackage{dsfont} 
\usepackage{mathtools} 

\begin{document}

\newcommand{\REF}[1]{\textcolor{red}{REF(#1)}}
\newcommand{\red}[1]{\textcolor{red}{#1}}
\newcommand{\TODO}[1]{\textcolor{red}{TODO #1}}

\newcommand{\bra}[1]{\left\langle #1 \right\vert }
\newcommand{\ket}[1]{\left\vert #1 \right\rangle }
\newcommand{\ev}[1]{\left\langle #1 \right\rangle }

\newcommand{\sx}{\sigma^x}
\newcommand{\sz}{\sigma^z}

\newcommand{\vecr}{\mathbf{r}}
\newcommand{\veck}{\mathbf{k}}
\newcommand{\pipi}{\left(\frac{\pi}{2}, \frac{\pi}{2}\right)}

\newcommand{\Cth}{C_3^h}
\newcommand{\Ctk}{C_3^k}
\newcommand{\Cf}{C_4}

\newcommand{\id}{\mathds{1}}

\title{False vacuum decay in a two-dimensional quantum spin system}

\author{Luka Pave\v{s}i\'{c}}
\email{luka.pavesic@unipd.it}
\affiliation{Dipartimento di Fisica e Astronomia “G. Galilei”, Universit\`a degli Studi di Padova, via Marzolo 8, I-35131 Padova, Italy}
\affiliation{Istituto Nazionale di Fisica Nucleare (INFN), Sezione di Padova, I-35131 Padova, Italy}
\author{Ian G. Moss}
\email{ian.moss@newcastle.ac.uk}
\affiliation{School of Mathematics, Statistics and Physics, Newcastle University, 
Newcastle Upon Tyne, NE1 7RU, UK}
\author{Simone Montangero}
\email{simone.montangero@unipd.it}
\affiliation{Dipartimento di Fisica e Astronomia “G. Galilei”, Universit\`a degli Studi di Padova, via Marzolo 8, I-35131 Padova, Italy}
\affiliation{Istituto Nazionale di Fisica Nucleare (INFN), Sezione di Padova, I-35131 Padova, Italy}

\begin{abstract}
False vacuum decay describes the relaxation of a metastable state through the nucleation and growth of bubbles of the stable phase. 
Despite describing a broad variety of phenomena across different fields, the quantum version of the nucleation theory has little experimental or numerical support. 
Testing its predictions is particularly important in two or more spatial dimensions, where bubble nucleation acquires its true geometrical nature. 
Here, we study false vacuum decay in the quantum Ising model in two dimensions.
Through tree tensor network simulations we extract the decay rate, the effective interface tension and the critical bubble size.
We compare them to new semi-classical field theory calculations, and find excellent agreement.
These results provide numerical evidence that the critical-bubble picture survives in an interacting quantum spin system in 2+1 dimensions.
\end{abstract}

\maketitle

Understanding dynamical properties of metastable states is of interest for a wide range of fields.
They typically appear after first-order phase transitions, and are thus ubiquitous at all scales:  in elementary particle theory it is believed that the Higgs field could be trapped in a metastable configuration\cite{Degrassi:2012ry,Buttazzo:2013uya}, and the same physics underpins the dynamics of phase separation in condensed matter~\cite{Langer1969_statistical, Binder1987, Bray1994}.
In the context of quantum dynamics, metastable states stand as examples of long-lived states that undergo slow thermalization, not due to the fragmentation of the Hilbert space, but rather because their decay is intrinsically non-perturbative~\cite{Yin2025}.
Understanding how long-lived states decay, and how to stabilise them, finds applications in quantum technologies, such as quantum memories and error correction~\cite{Terhal2015, Brown2016}. 

The canonical example of a many-body metastable state is the \emph{false vacuum}: the ground state of a local potential minimum, separated from the global minimum (\emph{true vacuum}) by a potential barrier. 
Its decay is described by \emph{critical bubble theory}: fluctuations -- either thermal or quantum -- drive the nucleation of domains in which the system locally converts to the stable phase. 
Such bubbles can either grow or collapse, depending on the competition between two contributions to their energy: the gain of converting the bulk of the droplet to the true vacuum, and the cost of the interface separating it from the metastable background. Since the bulk contribution grows faster with the bubble size than the interface contribution, there exists a threshold size beyond which the former becomes dominant.
Once a droplet larger than this threshold, \emph{the critical bubble}, is nucleated, its expansion is energetically favourable, eventually driving the whole system into the stable phase.

Here, we tackle the problem in the minimal lattice setting where quantum nucleation is driven by this geometric interplay: the quantum Ising model on a square lattice. 
Our goal is to test whether the semiclassical critical-bubble picture survives in an interacting quantum spin system in 2+1 dimensions, and to numerically determine its central quantities -- the decay rate, the interface tension, and the critical bubble size -- directly from simulations of many-body dynamics.

The theory of bubble nucleation was first developed by Langer~\cite{Langer:1967ax} for thermal phase transitions in classical Ising ferromagnets. 
Coleman introduced the concept of false vacuum decay and extended the theory to quantum phase transitions~\cite{Coleman:1977py,Callan:1977pt}, where the description requires an extra dimension, identified by an imaginary time coordinate. 
In this case, the critical bubble becomes a \emph{bubble instanton}. The nucleation event is now represented by a cross section of the instanton solution, and the rate of nucleation of bubbles $\gamma$ is exponentially suppressed by its action. 

\begin{figure*}
    \centering
    \includegraphics[width=\textwidth]{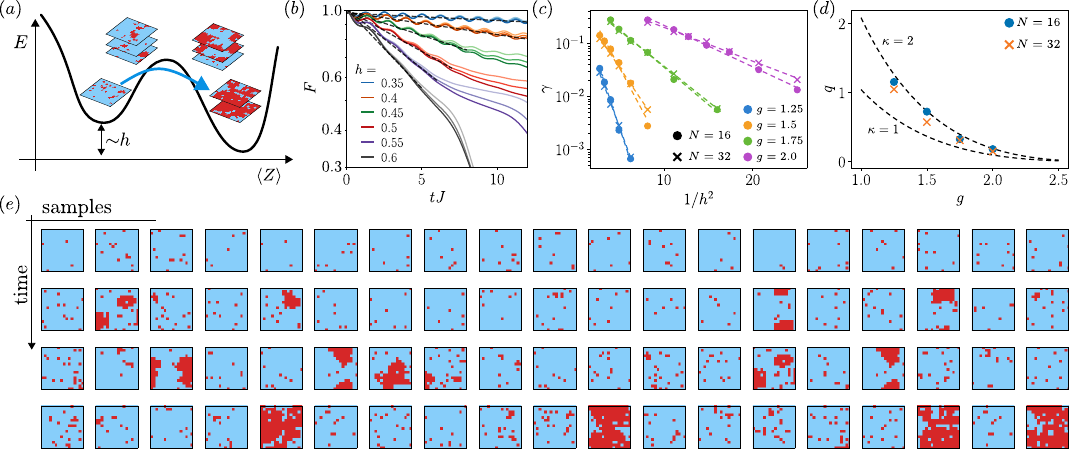}
    \caption{
    False vacuum decay in the 2D quantum Ising model.
    (a) A sketch of false vacuum decay in the 2D quantum Ising model. We study the relaxation of a weakly-correlated metastable state through the nucleation of growing true vacuum domains. 
    (b) Time evolution of $F(t)$ for $g=1.5$, $N=32$. Darker shades of the same color indicate calculations with increasing bond dimension; $\chi = 150$, $200$, $250$. Dashed black lines are fits to Eq.~\eqref{eq:decay_rate}.
    (c) Decay rate $\gamma$ vs $1/h^2$ across a set of $g$ and $N$.
    (d) Interface tension $q$ vs $g$ (points), compared to the analytical prediction (dashed lines).
    (e) Random samples of projective measurements in the computational basis at different times, showing the nucleation of true vacuum bubbles (red).
    }    
    \label{fig:decay_rate}
\end{figure*}

Despite the wide range of applications, the theory of false vacuum decay has little experimental support. 
The thermal version has been investigated, with various degrees of success, in liquid Helium~\cite{Schiffer1995,Balibar2002,Tian2023,Hindmarsh2024} and cold atomic gases~\cite{Zenesini2024, Cominotti2025}. For the latter case, numerical simulations have been extended into the zero temperature regime 
\cite{Fialko_2017,Billam:2021nbc,Jenkins:2023eez,Jenkins:2023npg}. 
In this quantum realm, the ordered phases of spin models contain both essential ingredients -- metastability and interface tension -- and are thus useful toy models for numerical studies of false vacuum decay.
The decay of metastable states in 1D spin chains was first numerically tested in Ref.~\onlinecite{Lagnese2021}; it was found to be consistent with semiclassical expectations of Ref.~\onlinecite{Rutkevich1999}.
Subsequent works studied the dynamics of nucleated bubbles~\cite{Sinha2021, Milsted2022, Johansen2025}, the role of lattice effects~\cite{Pomponio2022}, measurement~\cite{Maki2023}, and extensions beyond the Ising chain~\cite{Lencss2022, Pomponio2025}.
Progress has also come from programmable quantum simulators; dynamics of bubbles has been simulated on various platforms~\cite{Darbha2024_a, Darbha2024_b, Vodeb2025, Luo2025, Chao2026}.

However, the one-dimensional (1D) case is conceptually limited, because the size of the bubble interface does not depend on its size. 
Capturing the geometric nature of nucleation therefore requires going to at least two spatial dimensions. 
This is a major conceptual step. 
In 1D, much of the physics can be understood by appealing to a dual fermionic description; no such mapping is available in 2D. 
Analytical approaches are therefore necessarily approximate and rely on assumptions whose validity can be difficult to control. 
Moreover, the interface tension, a key ingredient of bubble-nucleation theory, is only known analytically in 1D.
New analytical expressions for the decay rate are therefore needed, together with numerical and experimental tests of their predictions.
Although recent work on two-dimensional spin systems has addressed the coarsening of domain walls and the resulting collective dynamics~\cite{Balducci2022, Krinitsin2025, pavesic2025, Manovitz2025}, the formation and growth of domains at special resonant points~\cite{Osterholz2025, Humar2026}, and impurity-seeded nucleation~\cite{Borla2026}, the spontaneous nucleation of critical bubbles remains unexplored.

We study this process in the quantum Ising model in two spatial dimensions.
Using tree tensor networks (TTN)~\cite{Tagliacozzo2009, Murg2010, Montangero_book_2026} we simulate the dynamics of the metastable false vacuum, extract the decay rate, the interface tension and the critical bubble size, and investigate the microscopic processes driving the bubble nucleation and growth. 
By comparing the numerical results to analytical field-theoretical calculations, we assess to what extent the semiclassical picture of nucleation survives in an interacting quantum spin system in $2+1$D, and present a quantitative picture of the interface contribution to the decay rate.

\emph{Model.}--
The minimal model for false vacuum decay in 2+1D is the quantum Ising model on a $N \times N$ square lattice:
\begin{equation}
    H = - J\sum_{\langle i j\rangle} Z_i Z_j - g \sum_iX_i - h \sum_i Z_i,
\end{equation}
where $X$ and $Z$ are Pauli matrices, and $\langle ij \rangle$ denotes a sum over the nearest neighbours with periodic boundary conditions. 
The interaction strength $J=1$ sets the energy scale, $g$ is the transverse and $h$ the longitudinal field. We use units in which the reduced Planck constant $\hbar=1$.

In the ordered phase ($g/J < 3.04$) the model has two degenerate ground states, with spins polarized in either $Z$ direction. 
The longitudinal field splits them in energy, turning the semi-classical potential into an asymmetric double well, as sketched in Fig.~\ref{fig:decay_rate}(a).  
The false vacuum is a low-energy state on the bottom of the metastable minimum. It is resonant with a continuum of highly excited states above the true vacuum, but transitioning to them requires flipping of an extensive number of spins. These matrix elements are thus $\sim (g/J)^N$, making the state's lifetime exponentially long~\cite{Yin2025}.  
The state relaxes through a non-perturbatively long chain of virtual processes which generate growing domains of positive magnetization.
It should be noted that this picture relies on the presence of a dense continuum of states in resonance with the initial state, which requires a large-enough system.
We find considerable finite-size effects for systems smaller than $N=16$, and use $N=16$ and $N=32$ for the results presented here. 

We use the density matrix renormalization group algorithm~\cite{Schollwock2011, Cirac2021} to find the ground state at some $(g, h)$, and simulate the evolution after a sudden quench $h \rightarrow -h$~\cite{Lagnese2021}. 
This procedure prepares a low-energy state in the metastable manifold, with the slight asymmetry between the excitations above the true and false vacua\footnote{The amplitude of vacuum fluctuations in leading order of $g/J$ is $g^2/(8J \pm 2h)$, with $-$ for the true and $+$ for the false vacuum.} providing the fluctuations to trigger the decay.

\emph{Decay rate.}--
We use rescaled mean magnetization~\cite{Lagnese2021}
\begin{equation}
    F(t) = \frac{Z(t)+Z(0)}{2Z(0)},
\end{equation}
with $Z(t) = \langle \psi(t) \vert Z \vert \psi(t)\rangle$, as a measure of the false vacuum fraction. 
Because the timescales associated with the decay are exponentially sensitive to the model parameters, some care is required to find the parameter range where the targeted dynamics is both clearly visible, but also quick enough to be accessible with tensor network simulations. In practice, intermediate $g \sim 1-2$ and $h \sim 0.2 - 0.5$, lead to substantial decay of $F$ on the timescale of $~\sim 10J$.

If the observed dynamics is described by the critical bubble theory, $F$ is to decay exponentially as $F(t) \propto e^{-\gamma t}$.
In the field-theoretic framework under quite general assumptions (see the Supplementary material~\cite{SM}), the decay rate $\gamma$ for the quantum Ising model in 2D takes the form
\begin{equation}
    \label{eq:decay_rate}
    \gamma = c(g)\,e^{-q(g)/h^2},
\end{equation}
where $q(g)$ is the interface tension, and $h$ is proportional to the energy difference between the true and the false vacua.
Our aim is to confirm the $h$ dependence of the decay rate, and to determine the functional form of the interface tension. For the tension term, we discard the $g$ dependence of the prefactor $c(g)$ as it is dominated by the exponential dependence ~\cite{SM}. 

The evolution of $F$ is shown in Fig.~\ref{fig:decay_rate}(b).
We find an overall exponential decay throughout the explored parameter range. 
The curves also exhibit small, but persistent oscillations at a frequency of order $J$, caused by repeated creation and annihilation of low-energy excitations. This is expected after quenches at $g \ll J$, where the elementary excitations are confined~\cite{Balducci2022, pavesic2025, Krinitsin2025, Tindall2024}. 
In the context of bubble nucleation however, excitations that collapse instead of growing can be interpreted as sub-critical bubbles. 
Extracting the biggest such domain sets a lower bound for the size of the critical bubble.
We return to this problem later.

The oscillations are washed away at larger $h$ (and with increasing $g$, see SM~\cite{SM}), which is due to two factors.
First, at larger $g$ and $h$ the critical bubble size becomes comparable to the lattice size, so there are no more sub-critical bubbles. 
Second, increasing $g$ weakens the kinetic constraints that bind them, allowing the excitations to disperse and interact. 
The onset of this behaviour might be related to the coarsening transition studied in Ref.~\onlinecite{Krinitsin2025} at $h=0$. However, determining its critical point at finite $h$ is beyond the scope of this work.

We extract the decay rate $\gamma$ by fitting $F(t)$ to an exponential decay.
The fits are shown as black dashed lines in Fig.~\ref{fig:decay_rate}(b).
The fitting time window is chosen case-by-case; we aim at the initial exponential decay, and exclude the data that is poorly converged in bond dimension and starts exhibiting additional behaviour.
The decay rate $\gamma$ is shown in Fig.~\ref{fig:decay_rate}(c).
We find good agreement with the expected scaling $1/h^2$, and overall good convergence in system size. The fit uncertainty is smaller than the markers.
Determining $\gamma(h)$ is one important result of our work; it stands as numerical confirmation that the critical bubble theory applies to our model, and that it can be used to describe the relaxation of metastable states in interacting quantum systems in 2+1D.

By fitting $\gamma(h)$ to Eq.~\eqref{eq:decay_rate} we determine the interface tension $q(g)$, shown in Fig.~\ref{fig:decay_rate}(d).
In the absence of a rigorous theory for $q$, we perform a simple extension of the analytical field theory results from 1D to the Ising model in 2D, and obtain a conjecture for the interface tension $q$. See SM~\cite{SM} for details.
The effective theory comes with a single free parameter $\kappa$ which corrects for the shape of the nucleated bubbles, taking values between $\kappa=1$ for square, and $\kappa=2$ for completely spherical bubbles. 
We show both predictions as dashed lines. The numerical data agrees best with $\kappa\approx 2$, suggesting that the bubbles are more circular than square.
This is curious: the shape minimising the interface enclosing a volume on a lattice is a square, so one could expect $\kappa \sim 1$. 
Similar behaviour was recently observed in a related study of bubble growth~\cite{Humar2026}.

\begin{figure*}
    \centering
    \includegraphics[width=\textwidth]{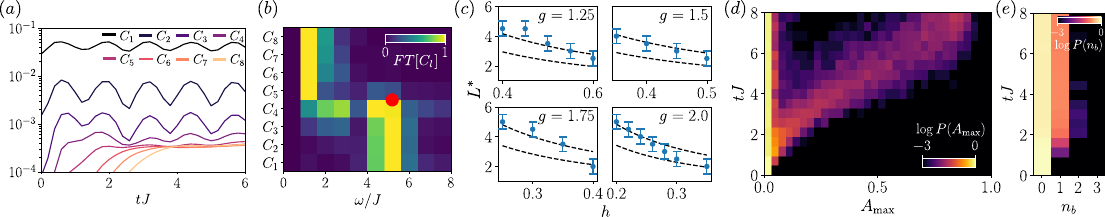}
    \caption{
    Critical bubbles and their growth.
    (a) Correlators $C_l$ for $l\in[1,8]$. $g=1.5$, $h=0.35$.
    (b) Fourier transform of $C_l$ shown in (a). The power spectrum is normalized for each $L$ for visual clarity. The size of the critical bubble is denoted with a red dot.
    (c) The extracted size of the critical bubble (points), compared to the analytical prediction for $\kappa=1$ (lower) and $\kappa=2$ (upper dashed line). Error bars are $\pm 1/2$ and come from the discreteness of available $C_L$. 
    (d) Distribution of the the area of the largest bubble $A_{\max}$ relative to the system size. Bin width is 8.
    (e) Distribution of the number of bubbles bigger than 10 spins per sample.  
    For (e,d): $g=1.5$, $h=0.4$, $\chi=250$. Statistics from $2000$ samples per time step.
    } 
    \label{fig:bubbles}
\end{figure*}

\emph{Critical bubbles.}--
Beyond the decay rate, the other central quantity of interest is the size of the critical bubble.
A similar question was recently addressed in Ref.~\onlinecite{Borla2026}, demonstrating the importance of bubble shape and the geometry of the underlying lattice. However, the effect of the surface tension renormalization and the broadening of domain walls at finite $g$ remains unexplored.

Directly extracting the bubbles from the entangled states is not straightforward: they take many arbitrary shapes, and measuring many-body correlators with tensor networks is prohibitively expensive. 
We thus restrict ourselves to a linear measure: define $\hat{n}_{ij} = (Z_{ij}-1)/2$, giving 1 if site $(i,j)$ is in the true vacuum and 0 otherwise, and measure linear cuts $C_L = (1-n_{ij}) n_{i+1,j} \dots n_{i+L,j} (1-n_{i+L+1,j})$ to detect a cross section of a true vacuum bubble of linear size $L$. 
An example of the evolution of $C_L$ is shown in Fig.~\ref{fig:bubbles}(a). See SM~\cite{SM} for an extended dataset.
We find that for small $L$, $C_L(t)$ oscillates, while above some threshold $L^*$ it monotonically grows. 
The oscillations signal that bubbles of this size are below the critical size and collapse, whereas the monotonic increase corresponds to growth. 
We thus conjecture $L^*$ as a measure of the critical bubble size.

To better quantify this distinction, we Fourier-transform $C_L$ into frequency space. The power spectrum $FT[C_L](\omega)$, shown in Fig.~\ref{fig:bubbles}(b), exhibits a clear peak at $\omega/J \sim 2\pi$ for small $L$. We define $L^*$ as $L$ where the peak suddenly vanishes. 
We extract it for a range of $g$, $h$, see Fig.~\ref{fig:bubbles}(c), and turn again to the comparison with effective field theory. It provides us with a prediction for the size of the nucleated critical bubble (instanton), shown as dashed lines for $\kappa=1$ (lower line) and $\kappa=2$ (upper). 
We again find surprisingly good agreement with the numerics for $\kappa$ closer to 2. 

It should be noted that this method of extracting $L^*$ should be interpreted with some care. 
The linear correlators $C_L$ provide a practical diagnostic of subcritical bubbles, but they are not a direct measurement of the full 2D droplet geometry. 
In particular, a linear domain of length $L$ can still be subcritical even when a more compact droplet with a comparable linear extent would grow. 
The disappearance of oscillations in $C_L$ therefore gives an effective critical length, which is likely an upper estimate of the minimal compact critical bubble. 
This also explains why the oscillation amplitude decreases smoothly with $L$, rather than disappearing at a perfectly sharp threshold. 

\emph{Bubble growth.}--
Finally, we ask how the nucleated bubbles grow. 
To this end, we sample projective measurements onto the computational basis, and extract bubble statistics from them. 
A set of random samples at four time steps (rows) is shown in Fig.~\ref{fig:decay_rate}(e). 
From each sample we extract the area of the biggest bubble $A_{\max}$ (ie. a connected domain of true vacuum), and plot the evolution of their distribution $P(A_{\max})(t)$ in Fig.~\ref{fig:bubbles}(d). 
These results are a further confirmation of the bubble nucleation picture: initially, the state only contains small fluctuations. As larger bubbles nucleate, the distribution develops two peaks. The majority is concentrated close to zero, corresponding to local excitations and collapsing sub-critical bubbles.
Meanwhile the second, smaller peak moves towards larger values of $A_{\max}$ as time progresses and the rare critical bubbles grow. 
At late times, the state is dominated by two types of structures: either sub-critical fluctuations, or macroscopically large true-vacuum domains. 
This does not mean that new bubbles are not nucleated -- sampling a freshly nucleated bubble is simply much less probable than observing one of the many already existing ones. 

A step beyond the theory of single-bubble nucleation is to ask how separate nucleation events are correlated. 
We probe this question by counting the number of bubbles present per sample: in Fig.~\ref{fig:bubbles}(e) we show the probability of finding $n_b$ bubbles bigger than 20 spins (for this parameter point $L^* \approx 4$, so the critical bubble area is $\sim 16$). 
We overwhelmingly find either zero or one such bubble per sample. 
A small peak at $n_b=2$ around $tJ \sim 2$ implies that some two-bubble configurations contributions are generated initially, but that these bubbles quickly merge. 
In the case of uncorrelated nucleation events, one would expect $P(n_b=2) \approx P(n_b=1)^2$. In our data we find at most $P(n_b=2)/ P(n_b=1)^2 = 0.2$, which may suggest that nucleation events are anti-correlated.
This assertion is admittedly inconclusive; further studies with much more extensive sampling, larger system sizes and better precision are required to thoroughly understand it. 

\emph{Discussion.}--
We studied the relaxation of a metastable false vacuum in the two-dimensional quantum Ising model, which provides a minimal lattice setting for quantum false vacuum decay.
In the explored regime, the false vacuum fraction displays an initial exponential decay with a rate consistent with semiclassical scaling expected for nucleation in 2D.
We extracted the effective interface tension and found it to be compatible with the instanton-based field-theory prediction.
We further identified the critical bubble size, again finding good agreement with the effective theory.
Taken together, these results provide numerical evidence that the critical-bubble picture survives in an interacting quantum spin system in $2+1$D.

This agreement is perhaps surprising.
The process studied here takes place on a lattice, at finite transverse field, and in a regime where the interface tension is strongly renormalised from its classical value.
The success of the semiclassical theory therefore suggests that its most robust ingredient is not the microscopic structure of the instanton, but the geometric competition between bulk energy gain and interface cost.
Microscopic details, quantum fluctuations, finite width of the interface, etc. appear mainly to renormalise the effective tension and bubble shape.
By contrast, the $1/h^2$ scaling of the decay rate and the approximate critical size are controlled by dimensionality and by the existence of a well-defined interface. 
These features are also expected to be present in the quantum field theory arena in which false vacuum decay was originally introduced. 
Indeed, the correspondence between nucleation in the Ising model and in the field theory stems from the reformulations of the Ising model in terms of quantum field theory that have been used previously to study critical phenomena~\cite{Langer:1967ax,Guenther:1979td,Lowe_1980}.

We see a natural direction for future work in developing more direct probes of bubble formation and growth, and interactions among them.
One possible route towards this goal would be to break translational invariance and deterministically induce nucleation in a chosen point.
This could be achieved by locally concentrating excess energy~\cite{pavesic2026}, engineering inhomogeneous quench protocols, or introducing impurities in a way that locally increases the nucleation rate.
Such protocols would make it possible to determine directly which bubbles grow and which collapse, and determine their critical size and shape.

Another direction would be to study nucleation beyond the critical bubble theory, whose central assumption is that nucleation occurs in a dilute regime where bubbles form rarely, grow in isolation, and do not interact.
This might be a good approximation for the decay of the Higgs field where the nucleation rate is minuscule, but need not hold in strongly interacting quantum systems. There, the relaxation may involve several bubbles nucleating within the same causal region, interacting, and coalescing, which the theory should encompass~\cite{Hawking1982, Lewicki:2019gmv}.
The ability to deterministically nucleate bubbles of controlled size, shape, and position in a quantum system would open the door to detailed studies of interacting regimes, and allow the microscopic dynamics of bubble growth and interaction to be probed directly.

Finally, the presented work can be interpreted as another in the series of results that establish quantum spin systems as an appropriate setting to study false vacuum decay and related field-theory phenomena, this time in 2+1D.
These can be naturally extended to implementations on quantum simulators, which promise access to regimes beyond the reach of classical numerics in the near future.
A particularly interesting feature of quantum simulators would be the ability to combine weak measurements and post-selection which would allow one to statistically follow the trajectory of an expanding bubble. Given the relevance of false vacuum decay to models of quantum cosmology~\cite{Coleman:1980aw,Linde:1986fe,Guth:2000ka}, where the bubble interiors and exteriors have different values of the cosmological constant, these simulations may offer a testbed for recent ideas on post-selection as an explanation of dark energy ~\cite{Anastopoulos:2024tsa,Lionas2026, Davies2026}.

\section*{Acknowledgements}
We are grateful to Lorenzo Maffi, Gianluca Lagnese, Iacopo Carusotto, Davide Rattacaso and Alex Jenkins for insightful discussions and comments on the manuscript. 
The tensor network calculations were performed with the \textsc{Quantum TEA} library~\cite{qtealeaves, qredtea, qredtea_benchmark}.

The work was supported by the European Union via
ICSC - Italian Research Center on HPC, Big Data and Quantum Computing (NextGenerationEU Project No. CN00000013), 
Horizon Europe program HORIZON-CL4-2022-QUANTUM-02-SGA via the project 101113690 (PASQuanS2.1), 
by the Italian Ministry of University and Research (MUR) via Quantum Frontiers (the Departments of Excellence 2023-2027),
by the World Class Research Infrastructure - Quantum Computing and
Simulation Center (QCSC) of Padova University, 
and by the Istituto Nazionale di Fisica Nucleare (INFN): iniziativa specifica IS-QUANTUM.
IGM is supported by the Science and Technology
Facilities Council of the UK, grant ST/X000567/1. 
The authors acknowledge computational resources of the INFN Padova HPC cluster, funded by the NextGenerationEU Terabit project on PNRR – Avviso n. 3264 ”per il Rafforzamento e creazione di Infrastrutture di Ricerca”, Missione 4, ”Istruzione e Ricerca” and of the Leonardo supercomputer hosted by CINECA (Italy) and the LEONARDO consortium.

\bibliography{bibliography}

\clearpage
\newpage
\onecolumngrid

\renewcommand{\theequation}{S\arabic{equation}}
\renewcommand{\figurename}{Supplementary Figure}
\setcounter{equation}{0}
\setcounter{figure}{0}     

\section*{Supplementary Material for: False vacuum decay in a two dimensional quantum spin system}
\FloatBarrier

\section*{Critical bubbles and instantons}

The application of critical droplet theory to ferromagnetic phase transitions in Ising models was pioneered by Langer \cite{Langer:1967ax}. Coleman's work extended this to a theory of false vacuum decay for application to quantum field theory, but the key ideas can be translated readily to spin systems. 
The fundamental concept is the bubble instanton, a solution to the field equations that uses an imaginary time coordinate. 
The nucleation rate $\gamma$ is related to the instanton action $B$,
\begin{equation}
    \gamma = c\,e^{-B}.  
\end{equation}
The pre-factor $c$ depends on the size of the system and fluctuations about the instanton solution. 
In order to apply Coleman's instanton theory to the quantum Ising model in two dimensions we make use of the fact that the quantum Ising model has a path integral formulation as a classical Ising model in three dimensions \cite{Suzuki:1976ldt}, in which we can identify the extra dimension as imaginary time \cite{Moss:2025fpg}. Bubble instantons are classical spin configurations that extremise the effective action $S_E$, which should include quantum corrections. In analogy with the thin-wall approximation in field theory, these configurations consist of a region of true vacuum spin values surrounded by a wall of spin flips.

On scales larger than the lattice separation, the wall can be approximated by a smooth two-dimensional surface ${\bf x}(u,v)$ in the three dimensions of space and imaginary time. The wall has an interface tension $\sigma$, that depends on how the surface is embedded in three dimensions, as represented by the surface unit normal ${\bf n}$. 
The total action is a sum of surface ($dA$) and volume ($dV$) terms, offset by a constant so that the action vanishes in the true vacuum:
\begin{equation}
    S_E[{\bf x}]=\int\sigma({\bf n})dA-2h\int dV.
\end{equation}
Note the scaling law of the area and volume terms,
\begin{equation}
    S_E[\alpha {\bf x}]=\alpha^2\int \sigma({\bf n})dA-2h\alpha^3\int dV
\end{equation}
The action is stationary with respect to variations of ${\bf x}$ at the instanton solution ${\bf x}_b$. A fortiori, $\partial S_E/\partial \alpha=0$, then setting $\alpha=1$ implies the surface and volume terms of the instanton solution are related. We can therefore eliminate the surface term to leave
\begin{equation}
    B\equiv S_E[{\bf x}_b]=h\int dV.
\end{equation}
The scaling law has a further consequence that instanton solutions ${\bf x}_b$ scale $\propto h^{-1}$, and so the volume
scales as $h^{-3}$, and
\begin{equation}
    B=\frac{q}{h^2},
\end{equation}
where the interface tension term $q$ depends only on $g$.
This equation applies to both field theory and spin systems, and is fundamental to the existence of an instanton description.

An analytic expression for the interface tension in the present case is not known. 
However, if the bubble instanton has approximate rotational symmetry in the spatial plane, then the interface tension depends primarily only on imaginary time $\tau$. The dimensionality of the system is effectively reduced, and we conjecture that the instanton radius $r(\tau)$ is related to the size of the instanton in the same way as in the 1D quantum Ising model~\cite{Avron1982,Zia1982,Rutkevich1999}. 
The radius of the instanton, in lattice spins, would be given by  
\begin{equation}
    r=\frac{\sqrt{\kappa}}{h}\left(1+\frac{g^2}{g_c^2}-2\frac{g}{g_c}\cosh\theta\right)^{1/2},
\label{eq:radius}
\end{equation}
where $\theta=2hJ\tau$, $g_c\approx3$ is the critical point of the quantum Ising model in 2D and $\kappa$ is a parameter correcting for non-circular bubble shape.
The instanton action is related to the volume as above,
\begin{equation}
    B = h\int dV= \int_0^{|\ln g/g_c|} \pi r^2d\theta=\frac{q}{h^2}.
\end{equation}
The result is a conjecture for the interface tension $q$,
\begin{equation}
    \label{eq:interface_tension}
    q=\kappa\pi\left(1+\frac{g^2}{g_c^2}\right)\left|\ln \frac{g}{g_c}\right|-\kappa\pi\left|1-\frac{g^2}{g_c^2}\right|.
\end{equation}
The radius of the bubble when it nucleates at $\tau=0$, from Eq.~\eqref{eq:radius}, is 
\begin{equation}
    \label{eq:critical_bubble_size}
    r^* = \sqrt{\kappa}(1-g/g_c)/h.
\end{equation}
We compare this to the linear measure of the critical bubble size: $L^* \sim 2 r^*$ 

Calculation of the prefactor $c$ in the nucleation rate requires evaluating an operator determinant. In $d$ dimensions,  part of the result comes from zero modes, $B^{(d+1)/2}$, and a rescaling of the rest implies an {\it approximate} dependence on the bubble radius 
$(r^*)^{-(d+1)}$ \cite{Coleman:1977py,Callan:1977pt},
\begin{equation}
c\approx A\left(\frac{B}{r^{*\,2}}\right)^{(d+1)/2}
\end{equation}
where $A$ is constant.
In particular, $c\propto h^{(d+1)(1-d/2)}$. In the $d=2$ case,
\begin{equation}
c\approx A\frac{(q/\kappa)^{3/2}}{(1-g/g_c)^3}.
\end{equation}
Notice that there is no dependence on $h$ or $\kappa$ ($q$ is proportional to $\kappa$, see Eq.~\eqref{eq:interface_tension}), which means that when extracting nucleation exponents from runs at different values of $h$ it is possible to treat $c$ as a constant pre-factor.

\section*{Tree tensor networks}

The numerical calculations presented in the paper rely on the framework of tensor network methods.
We use tree tensor networks (TTN)~\cite{Tagliacozzo2009, Murg2010, Kohn2020}, where the state is presented as a network of tensors arranged into a binary tree. The tensors in the bottom layer contain the physical degrees of freedom, whilst the layers in the upper layers act as auxiliaries which efficiently transfer long-range correlations across the network.

TTN are loopless, so they retain many nice properties of the more standard matrix product states (MPS): they are efficiently contractable, so observables can be easily extracted from the network. Additionally, all relevant algorithms for manipulating MPS can be transposed to TTN. 
We use the time-dependent variational principle (TDVP)~\cite{Haegeman2016, Bauernfeind2020} to evolve them in time. 
We use the single-site version of TDVP, where each tensor is updated separately. This version of the algorithm cannot grow the bond dimension during the evolution, so our initial states start at the maximal bond dimension. This introduces some overhead in the initial steps, but we find that this approach performs best for evolving large systems to long times.
We use the same time step for all simulations: $dt = 0.01/J$.

TTN is a linear network, so the 2D lattice has to be mapped down to one dimension, using some space-filling curve.
This transforms a local model into one with effective long-range interactions, and thus the state typically develops long-range correlations (in the metric of the network).
This makes the TTN simulations less efficient, and requires to work at larger bond dimension. 
Choosing a good mapping (ie. one that minimizes the amount of long-range connections) is an important aspect of simulating 2D systems with MPS or TTN~\cite{Cataldi2021, Bellwood2025, Scardicchio2026}.
For the simulations on the largest lattices ($32 \times 32$ and $24 \times 24$), we used the optimal mappings provided in Ref.~\cite{Scardicchio2026}. For $16 \times 16$, we find the Hilbert curve proposed in Ref.~\cite{Cataldi2021} is sufficient.

\section*{The effect of the system size}

In a closed quantum system where the total energy is conserved, the dynamics of the initial state is determined by the manifold of states which are close in energy to the initial state. 
In our setting, this strongly depends on the system size: the energy of the false vacuum is $\sim h ~N^2$. 
Furthermore, the semi-classical theory expects that the manifold to where the state decays is an approximately featureless continuum.
The system has to be large enough to ensure that we are close to this approximation.

\begin{figure}
    \centering
    \includegraphics[width=0.4\columnwidth]{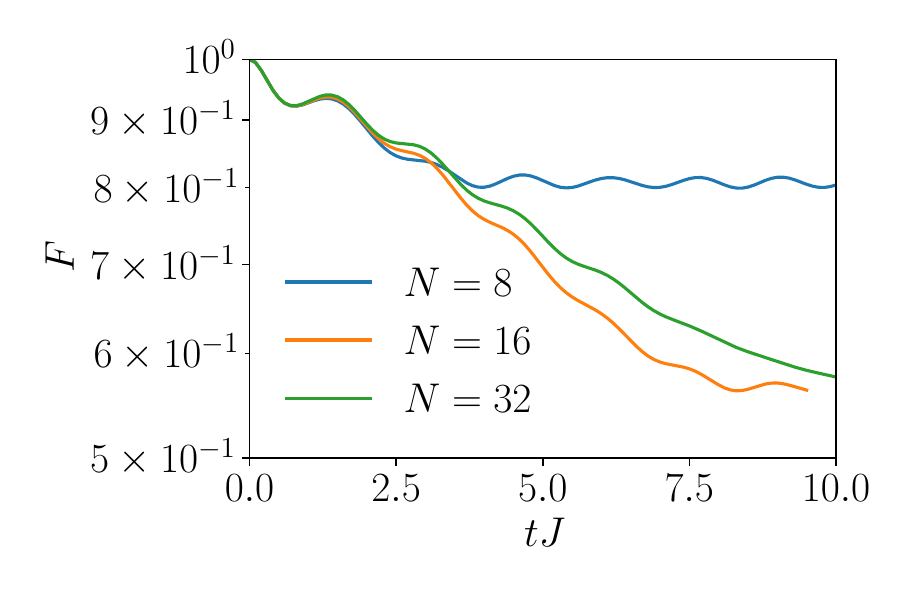}
    \caption{
    Evolution of $F$ for different $N$ at $g=1.5$, $h=0.5$; $\chi=250$.
    }
    \label{fig:SM:N}
\end{figure}

In Fig.~\ref{fig:SM:N} we show the evolution of $F$ for $g=1.5$, $h=0.45$ for $N = 8$, $16$, $32$.
We find that the initial decay rate does not depend on $N$. However, the late-time behaviour, in particular how far $F$ decays, is strongly $N$-dependent.
This is essentially because the growth of the bubbles is naturally cut off when the bubbles reach a size comparable to the system size.

\section*{Full dataset for $F$ and $C_L$}

\begin{figure}
    \centering
    \includegraphics[width=\columnwidth]{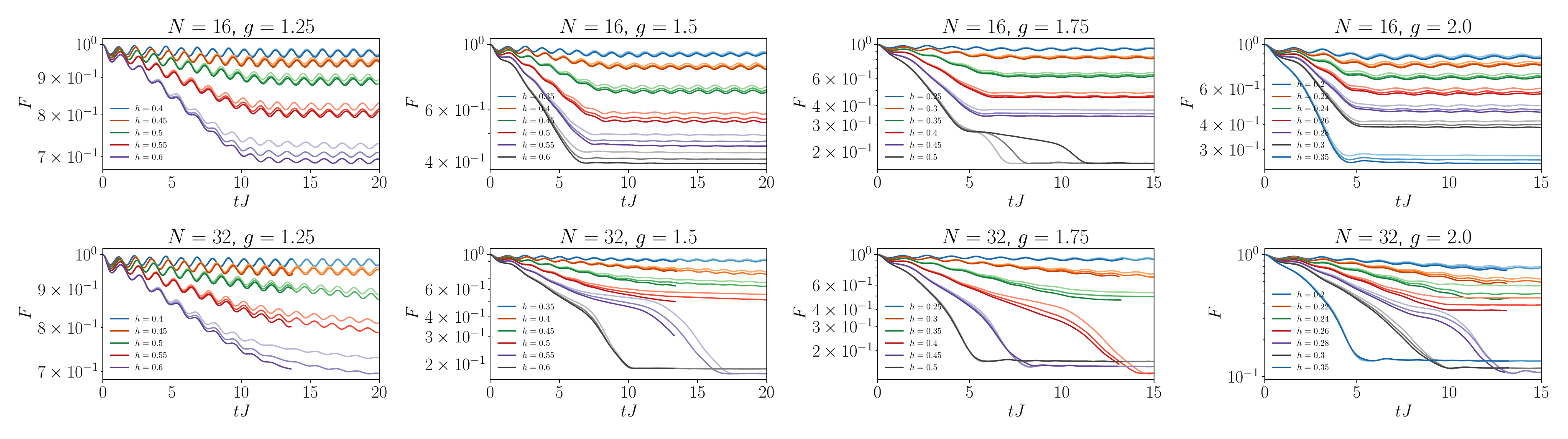}
    \caption{
    Evolution of $F$ for of $N$, $g$, $h$. Darker shades of the same color correspond to calculations with larger bond dimension, $\chi = 150$, $200$, $250$.
    }    
    \label{fig:SM:F}
\end{figure}

\begin{figure}
    \centering
    \includegraphics[width=\columnwidth]{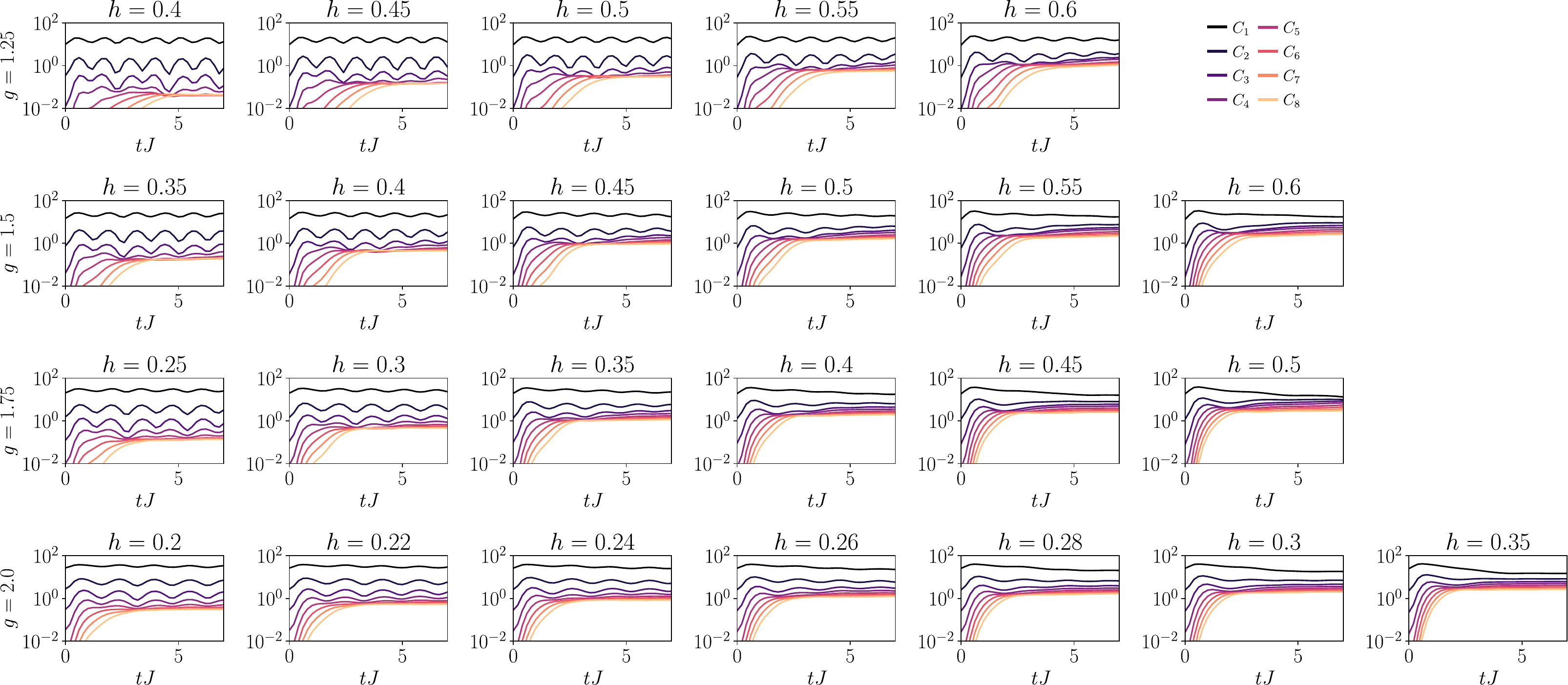}
    \caption{
    Evolution of $C_L$ for a set of $g$, $h$; $N=16$, $\chi=250$. 
    }    
    \label{fig:SM:C}
\end{figure}

\begin{figure}
    \centering
    \includegraphics[width=\columnwidth]{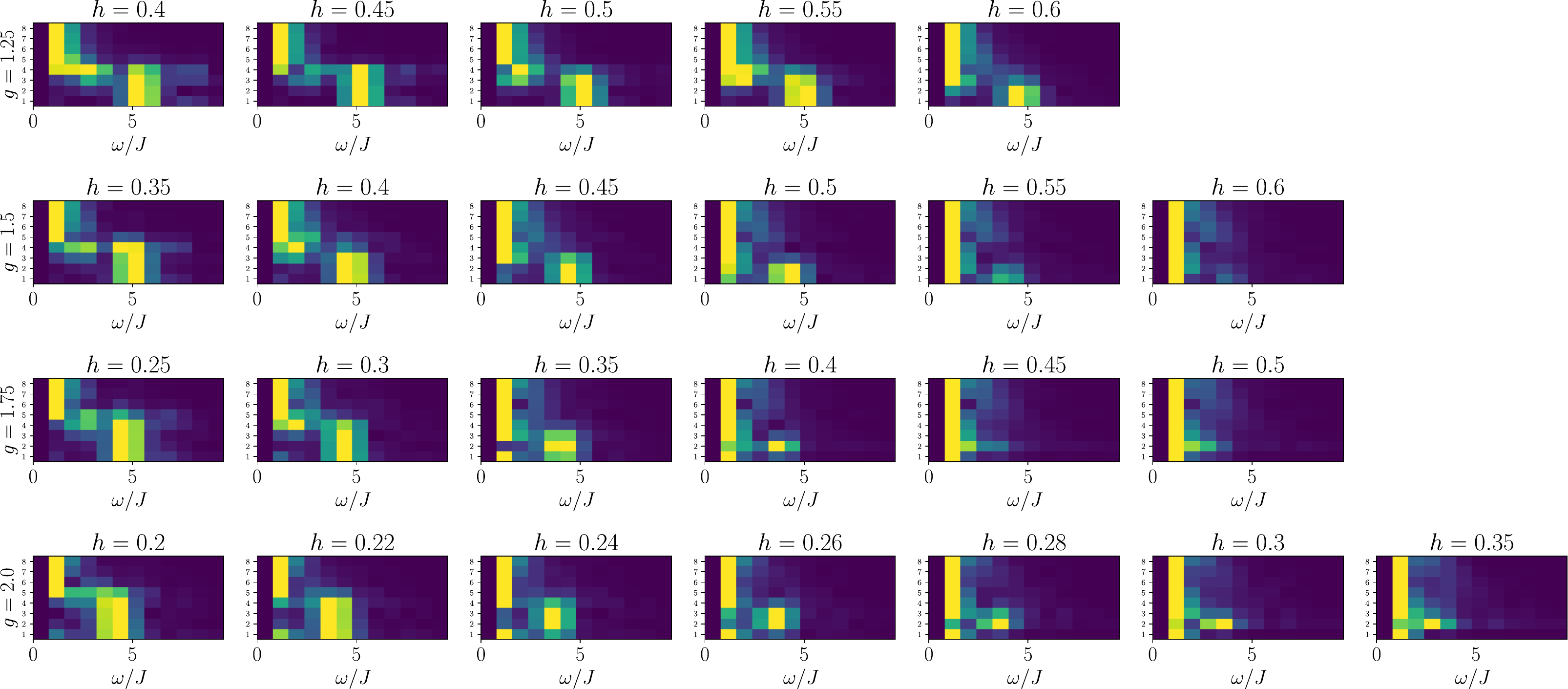}
    \caption{
    $\mathrm{FT}[C_L]$ for a set of $g$, $h$; $N=16$, $\chi=250$. 
    }    
    \label{fig:SM:FTC}
\end{figure}

Here we present the complete dataset across varying $g$, $h$.
Fig.~\ref{fig:SM:F} shows the evolution of $F(t)$, from which we extract the decay rates.
Fig.~\ref{fig:SM:C} shows the evolution of the correlators $C_L(t)$, while Fig.~\ref{fig:SM:FTC} shows the Fourier transformed data.

\end{document}